\documentclass[aps, apl]{revtex4-1}
\usepackage{graphicx}% Include figure files
\usepackage{subfigure}
\usepackage{bm}% bold math
\usepackage{color}

\usepackage{amsmath}
\usepackage{amssymb}
\usepackage{enumerate}
\usepackage{bm}

\usepackage[toc]{appendix}

\usepackage{lineno}
\usepackage{ulem}
\newcommand{\be}{\begin{equation}}
\newcommand{\ee}{\end{equation}}

\begin{document}

\title{A computational mean-field model of interacting non-collinear classical spins}
\author{O. Hovorka$^{1}$, T. Sluckin$^{2}$}
\affiliation{$^1$Engineering and Physical Sciences, University of Southampton, SO17 1BJ, Southampton, UK}
\affiliation{$^2$Mathematical Sciences, University of Southampton, SO17 1BJ, Southampton, UK}

\begin{abstract}
\noindent Mean-field approximation is often used to explore the qualitative behaviour of phase transitions in classical spin models before employing computationally costly methods such as the Monte-Carlo techniques. We implement a `lattice site-resolved' mean-field spin model that allows efficient simulation of phase transitions between phases of complex magnetic domains, such as magnetic helices, skyrmions, or states with canted spins. The framework is useful as a complementary approach for pre-screening the qualitative features of phase diagrams in complex magnets. 

\end{abstract}

%\pacs{75.10.Hk,75.20.-g,75.50.Ss,75.60.Jk,75.78.Jp}
%\linenumbers
\date{\today}
\maketitle

%%%%%%%%%%%%%%%%%%%%%%%%%%%%%%%%%%%%%%%%
%%%%%%%%%%%%%%%%%%%%%%%%%%%%%%%%%%%%%%%%
%%%%%%%%%%%%%%%%%%%%%%%%%%%%%%%%%%%%%%%%
\section{Introduction}

\noindent Classical spin models combined with Monte-Carlo simulations are often the method of choice for studying phase transitions \cite{landau2014guide}. However, high-quality Monte-Carlo simulations of phase diagrams are computationally costly. For this reason, it is often useful to begin the analysis by assuming a mean-field approximation (MF) based on neglecting the collective effects of thermal fluctuations, which allows accessing qualitative features of phase diagrams more efficiently. Although the MF models, in general, fail to predict the critical temperatures and critical exponents, they are nevertheless suitable for exploring the nature of existing thermodynamic phases of the system under study.\\

\noindent The simplest derivation of the MF approximation assumes translational invariance along the spin-lattice, which is suitable for a qualitative description of phase transitions between uniform phases such as the paramagnetic-to-ferromagnetic transition \cite{goldenfeld2018lectures}. However, description of more complex phases with non-collinear spins such as, for example, the helimagnetic order, magnetic bubbles or skyrmions, requires deriving MF approximations without assuming the translational invariance. In this work, we derive such a self-consistent lattice site-resolved MF model using a classical Heisenberg spins system (O(3)-model) with Dzyaloshinskii-Moriya interaction (DMI). We then highlight the computational algorithm that can be used for computing the mean-field spin configuration at a given magnetic field and temperature in an efficient manner and present some examples of calculations of complex spin textures, such as skyrmion lattices at finite temperatures.

%%%%%%%%%%%%%%%%%%%%%%%%%%%%%%%%%%%%%%%%
%%%%%%%%%%%%%%%%%%%%%%%%%%%%%%%%%%%%%%%%
%%%%%%%%%%%%%%%%%%%%%%%%%%%%%%%%%%%%%%%%
\section{Summary of the computational model}

\noindent Although the techniques used below to derive the mean-field formalism apply to a general class of classical spin Hamiltonians, for specificity we consider a Heisenberg spin system with DMI \cite{lancaster_skyrmions_2019}:
\be\label{spinham}
\begin{split}
\mathcal{H} =
-\frac{1}{2}\sum_{ij}J_{ij}\mathbf{s}_i\cdot\mathbf{s}_j
-\frac{1}{2}\sum_{ij}\mathbf{D}_{ij}\cdot(\mathbf{s}_i\times\mathbf{s}_j)
-\mu\sum_i\mathbf{s}_i\cdot\mathbf{B}_i,
\end{split}
\ee
where the individual terms represent the ferromagnetic exchange interaction energy, DMI energy, and a random magnetic field energy term. Uniform magnetic fields can be set by replacing $\mathbf{B}_i=\mathbf{B}$ for all $i$. The spin variables are unit vectors $\mathbf{s}_i=\bm{\mu}_i/\mu$, $i = 1,\dots, N$, where $\bm{\mu}_i$ is the magnetic moment associated with the spin $i$ and $\mu=|\bm{\mu}_i|$ is its magnitude.

%%%%%
\subsection{Mean-field model}
\noindent The `lattice site-resolved' mean-field model corresponding to Hamiltonian in Eq. \eqref{spinham} can be derived by using the standard field-theoretic arguments as outlined in Appendix \ref{appendix:app1}, which gives the mean-field energy as:
\be\label{spinhammf}
\begin{split}
\mathcal{H_\mathrm{MF}} =
-\frac{1}{2}\sum_{ij}J_{ij}\mathbf{\tilde m}_i\cdot\mathbf{\tilde m}_j
-\frac{1}{2}\sum_{ij}\mathbf{D}_{ij}\cdot(\mathbf{\tilde m}_i\times\mathbf{\tilde m}_j)
-\mu\sum_i\mathbf{\tilde m}_i\cdot\mathbf{B}_i,
\end{split}
\ee
where $\mathbf{\tilde m}_i$  represent mean-field spin vector moments with temperature dependent magnitude:
\be\label{lanf}
\mathbf{\tilde m}_i = \mathcal{L}\left(\beta\mu|\mathbf{\tilde B}_i^\mathrm{e}|\right)
\frac{\mathbf{\tilde B}_i^\mathrm{e}}{|\mathbf{\tilde B}_i^\mathrm{e}|}.
\ee
Here $\mathcal{L}(x) = \coth x - x^{-1}$ is the Langevin function, and $\beta = (k_BT)^{-1}$ with $k_B$ being the Boltzmann constant and $T$ the temperature. The vector $\mathbf{\tilde B}_i^\mathrm{e}$ represents the effective field acting on the moment $\mathbf{\tilde m}_i$, and $|\mathbf{\tilde B}_i^\mathrm{e}|$ is the magnitude of $\mathbf{\tilde B}_i^\mathrm{e}$. The expression for the effective field can be obtained by calculating the variational derivative of Eq. \eqref{spinhammf}, i.e. $\mu\mathbf{\tilde B}_i^\mathrm{e}=-\delta\mathcal{H_\mathrm{MF}}/\delta\mathbf{\tilde m}_i$, which gives:
\be\label{beff}
\begin{split}
\mu\mathbf{\tilde B}_i^\mathrm{e}
=J_{ij}\mathbf{\tilde m}_j
- \mathbf{D}_{ij}\times\mathbf{\tilde m}_j
+ \mu\mathbf{B}_i
+ \mu\mathbf{B}.
\end{split}
\ee
According to Eq. \eqref{lanf}, the MF spin moment $\mathbf{\tilde m}_i$ depends through $\mathbf{\tilde B}_i^\mathrm{e}$ on the couplings $J_{ij}$ and $\mathbf{D}_{ij}$, and the magnetic field $\mathbf{B}_i$, and also on temperature $T$.
It is worth pointing out that by taking the small angle approximation and zero temperature limit, the model above reduces to the standard micromagnetic energy \cite{bertotti1998hysteresis}.

%%%%%
\subsection{Iterative solution method}\label{sec:itersol}

\noindent Eqs. \eqref{lanf} and \eqref{beff} represent a system of $N$ coupled nonlinear equations, which can be solved by the following iterative technique based on:

\begin{enumerate}

\item[(i)] Initialising the system in a well-defined state, such as in the spin-aligned state obtained at high `saturating' external field $\mathbf{B}_i=\mathbf{B}$ at any temperature $T$, or in a random state obtained in the `paramagnetic' phase at high temperature $T$ and $\mathbf{B}=\mathbf{0}$.

\item[(ii)] Picking a spin at a lattice site $k$ and using Eq. \eqref{beff} to calculate the corresponding effective field vector $\mathbf{\tilde B}_k^\mathrm{e}$.

\item[(iii)] Calculating a new MF spin moment $\mathbf{\tilde m}_k'$ by using Eq. \eqref{lanf} and updating the previous moment $\mathbf{\tilde m}_k$ incrementally to $\mathbf{\tilde m}_k+\lambda(\mathbf{\tilde m}_k'-\mathbf{\tilde m}_k)$, where $\lambda$ is a convergence factor. Our typical choice $\lambda=0.4-0.6$ leads to good convergence rates most of the time.

\item[(iv)] Repeating the steps (ii) and (iii) for all MF spin moments $\mathbf{\tilde m}_k$, $k=1,\dots,N$, available in the lattice.

\item[(v)] Iterating between (ii)-(iv) until every lattice site $i$ has been repeatedly visited a sufficient number of times to reduce any differences between the new and previous mean-field spin configurations $\mathbf{\tilde m}_k$ to remain within the specified tolerance:
\be
\mathtt{tol}=\left(\sum_l\sum_k\left(\mathbf{\tilde m}_k^{(l)}-\mathbf{\tilde m}_k^{(l-1)}\right)^2\right)^{1/2}
\ee
 where $l$ specifies the iteration index. In our simulations, setting $\mathtt{tol}=10^{-5}$ resulted in a very good accuracy of the computed MF spin patterns.

\end{enumerate}

\noindent The algorithm above allows simulating the magnetisation curves at a given temperature. It allows computing the zero-field or field-cooled mean-field spin patterns. It can also be used to explore the metastable phase diagrams produced by arbitrarily chosen external field and temperature protocols, and thereby inform the computationally more involved Monte-Carlo methods.

%%%%%%%%%%%%%%%%%%%%%%%%%%%%%%%%%%%%%%%%
%%%%%%%%%%%%%%%%%%%%%%%%%%%%%%%%%%%%%%%%
%%%%%%%%%%%%%%%%%%%%%%%%%%%%%%%%%%%%%%%%
\section{Examples of simulations}

\begin{figure}[b]
\includegraphics[width=9cm]{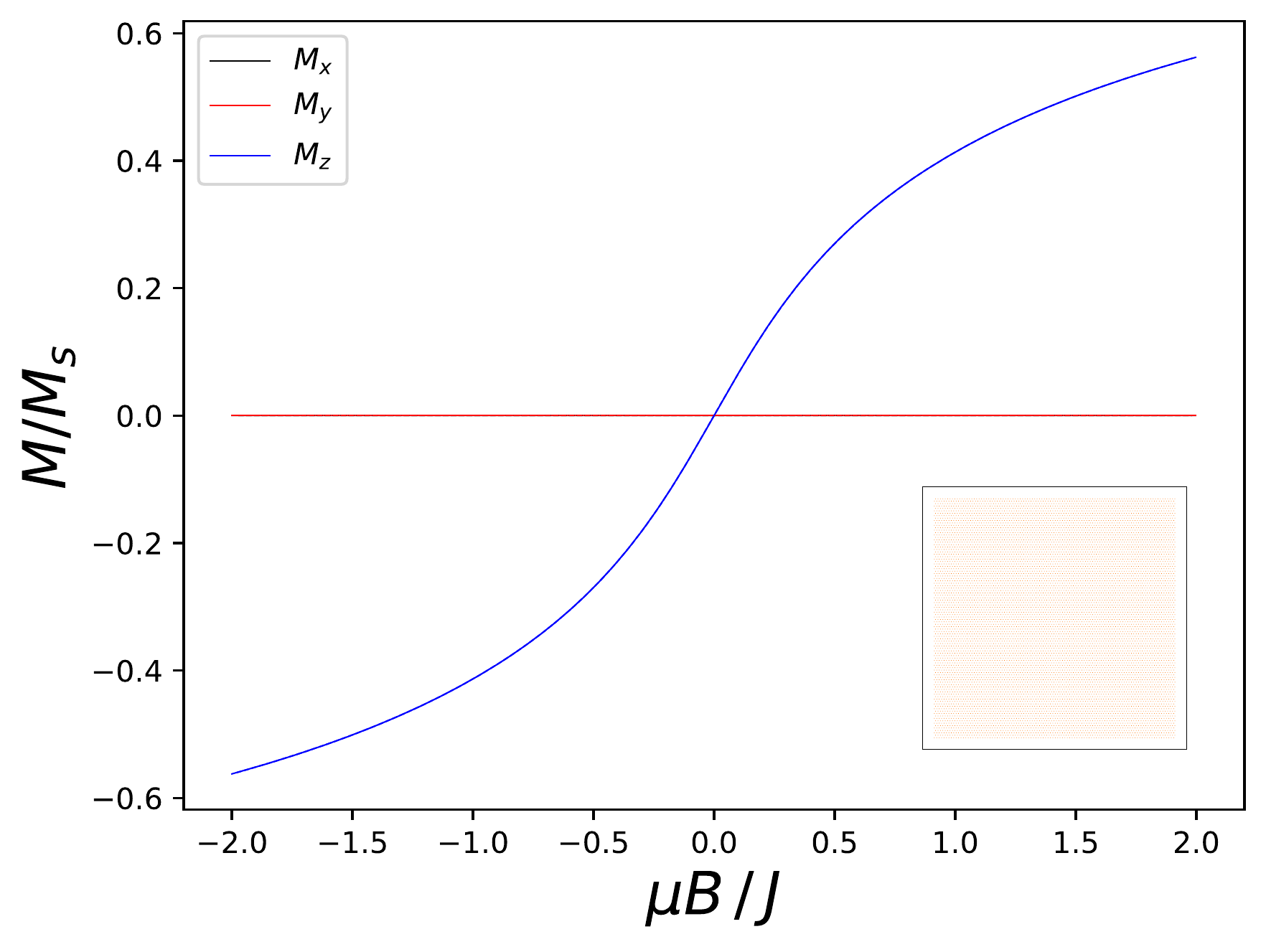}
\caption{Average mean-field spin moment vs. magnetic field at temperature $\beta=0.8$ in the paramagnetic regime. The inset shows a snapshot of a typical uniform mean-field spin configuration taken at a positive field $\mathbf{B} = B\mathbf{\hat z}$ with all spins pointing along the field direction. The remaining simulation parameters were $J_{ij}=J=D_{ij}=D=0.5$, all couplings restricted to the nearest neighbour spins, $N=100\times100$ square lattice with periodic boundary conditions, $\lambda=0.5$, and $\mathtt{tol}=10^{-5}$.}
\label{fig:hl1}
\end{figure}

\begin{figure}[t]
\includegraphics[width=9cm]{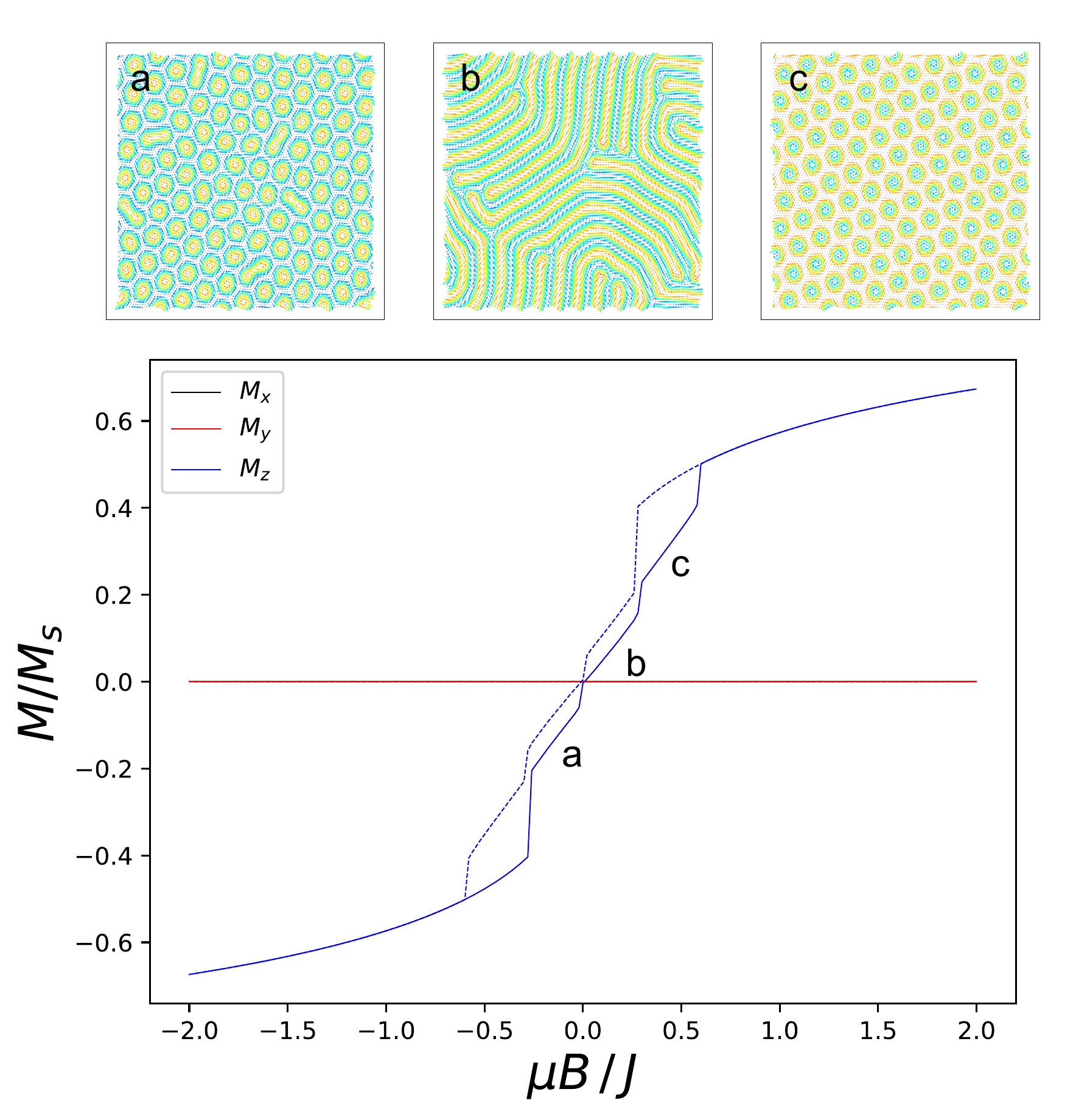}
\caption{Average mean-field spin moment vs. magnetic field at temperature $\beta=1.0$ below the phase transition point from a paramagnetic phase, displaying hysteresis behaviour. The insets \textbf{a}, \textbf{b}, and \textbf{c} show snapshots of the skyrmion lattice and helical mean-field spin configurations taken in the highlighted regions \textbf{a}, \textbf{b}, and \textbf{c} along the hysteresis curve. The remaining simulation parameters were $J_{ij}=J=D_{ij}=D=0.5$, all couplings restricted to the nearest neighbour spins, $N=100\times100$ square lattice with periodic boundary conditions, $\lambda=0.5$, and $\mathtt{tol}=10^{-5}$.}
\label{fig:hl2}
\end{figure}

\noindent Fig. \ref{fig:hl1} shows the magnetic field-dependence of the average mean-field spin state calculated as $\mathbf{M}=N^{-1}\sum_i\mathbf{\tilde m}_i$ in the paramagnetic temperature regime, showing a typical Langevin curve-like behaviour. The mean-field spins in the lattice always point along the magnetic field direction as shown in the example in the inset. When the temperature is lowered ($\beta$ increases) across the phase transition point hysteresis behaviour appears, which is characterised by complex non-collinear mean-field spin configurations shown in Fig. \ref{fig:hl2}. Specifically, the inset shows skyrmion lattices and helical structures appearing in the respective magnetic field regions along the hysteresis curve shown in Fig. \ref{fig:hl2}.

%%%%%%%%%%%%%%%%%%%%%%%%%%%%%%%%%%%%%%%%
%%%%%%%%%%%%%%%%%%%%%%%%%%%%%%%%%%%%%%%%
%%%%%%%%%%%%%%%%%%%%%%%%%%%%%%%%%%%%%%%%
\section{Conclusions}

\noindent The classical MF spin model developed in this work allows simulating complex spin textures at finite temperatures. Since the model does not rely on the continuum approximation requiring the small-angle assumption between neighbouring spins, it allows simulations of highly non-collinear spin configurations, including chiral and antiferromagnetic-like systems. Moreover, since the magnitude of the MF spin moments is allowed to vary as a function of the local effective field, Bloch point singularities often emerging in continuum theories can be naturally modelled as well. 

%%%%%%%%%%%%%%%%%%%%%%%%%%%%%%%%%%%%

% Bibliography - using database
\bibliography{meanfield}

%%%%%%%%%%%%%%%%%%%%%%%%%%%%%%%%%%%%%%%%
%%%%%%%%%%%%%%%%%%%%%%%%%%%%%%%%%%%%%%%%
%%%%%%%%%%%%%%%%%%%%%%%%%%%%%%%%%%%%%%%%
%%%%%%%%%%%%%%%%%%%%%%%%%%%%%%%%%%%%%%%%
%%%%%%%%%%%%%%%%%%%%%%%%%%%%%%%%%%%%%%%%
%%%%%%%%%%%%%%%%%%%%%%%%%%%%%%%%%%%%%%%%
%%%%%%%%%%%%%%%%%%%%%%%%%%%%%%%%%%%%%%%%
%%%%%%%%%%%%%%%%%%%%%%%%%%%%%%%%%%%%%%%%
%%%%%%%%%%%%%%%%%%%%%%%%%%%%%%%%%%%%%%%%
%%%%%%%%%%%%%%%%%%%%%%%%%%%%%%%%%%%%%%%%
%%%%%%%%%%%%%%%%%%%%%%%%%%%%%%%%%%%%%%%%
%%%%%%%%%%%%%%%%%%%%%%%%%%%%%%%%%%%%%%%%

\vfill
\pagebreak

\appendix

%%%%%%%%%%%%%%%%%%%%%%%%%%%%%%%%%%%%%%%%
%%%%%%%%%%%%%%%%%%%%%%%%%%%%%%%%%%%%%%%%
\section{Derivation of the mean-field model using field-theoretic approach}\label{appendix:app1}

\noindent The simplest way to obtain the mean-field (MF) model outlined by Eqs. \eqref{lanf}-\eqref{beff} in the main text is by expressing the spin fluctuations around the mean as $\mathbf{s}_i = \mathbf{\tilde m}_i+\delta\mathbf{s}_i$, i.e. the spin $\mathbf{s}_i$ fluctuates around the mean $\mathbf{\tilde m}_i$ by amount $\delta\mathbf{s}_i$ \cite{goldenfeld2018lectures}. Inserting this expression into Eq. \eqref{spinham}, arranging and neglecting the terms with $\delta\mathbf{s}_i$ beyond the first order allows expressing Eq. \eqref{spinham} in simpler form with all energy terms containing only a linear dependence on the spin variables $\mathbf{s}_i$. This approximation allows evaluating the underlying canonical partition function, from which Eq. \eqref{lanf}-\eqref{beff} can be obtained directly by using the relevant thermodynamic relations \cite{goldenfeld2018lectures}.\\

\noindent In this work, we follow a different derivation based on field-theoretic arguments. This approach allows obtaining the MF theory as the first-order term in the systematic expansion of the partition function. It has the potential advantage in that it allows deriving higher-order corrections by using the field-theoretic techniques.

\subsection{Spin Hamiltonian in the matrix form}

\noindent The effective spin Hamiltonian defined in Eq. \eqref{spinham} can be rewritten in matrix form:
\be\label{ham_matrix1}
\begin{split}
\mathcal{H} =
-\frac{1}{2}\mathbf{s}_i\cdot\mathbb{J}_{ij}\cdot\mathbf{s}_j
-\frac{1}{2}\mathbf{s}_i\cdot\mathbb{D}_{ij}\cdot\mathbf{s}_j
-\mu \mathbf{B}_i\cdot\mathbf{s}_i.
\end{split}
\ee
The first two terms describe the exchange and Dzyaloshinskii-Moriya interaction (DMI) energies, and the last term is the random field energy. The Einstein summation convention over the repeated indices is assumed. The indices $i, j = 1, \dots, N$ refer to the individual spins on the lattice. We introduced block matrices $\mathbb{J}_{ij}\rightarrow J_{ij}^{kl}=J_{ij}\delta_{kl}$ and $\mathbb{D}_{ij}\rightarrow D_{ij}^{kl} = D_{ij}^m\epsilon_{mkl}$, where $J_{ij}$ and $D_{ij}^m$ are non-zero for neighbouring spins and zero otherwise. The superscript indices $k, l, m = 1, 2, 3$ correspond to matrix elements multiplying the $x, y, z$ coordinates of the underlying vector moments of spins. The $\delta_{ij}$ is the Kronecker delta and $\epsilon_{klm}$ is the Levi-Civita symbol.
Introducing $\mathbb{I} = \mathbb{J} + \mathbb{D}$ as a combined interaction matrix, Eq. \eqref{ham_matrix1} can be expressed as:
\be\label{ham_matrix2}
\begin{split}
\mathcal{H} =
-\frac{1}{2}\mathbf{s}_i\cdot\mathbb{I}_{ij}\cdot\mathbf{s}_j
-\mu \mathbf{B}_i\cdot\mathbf{s}_i.
\end{split}
\ee
It is straightforward to check that $\mathbb{I}$ is a square matrix of size $3N\times 3N$, symmetric and invertible.

%%%%%%%%%%%%%%%%%%%%%%%%%%%%%%%%%%%%%%%%

\subsection{Partition function and Action}

\noindent The partition function assuming Eq. \eqref{ham_matrix2} is:
\be\label{pf0}
\begin{split}
\mathcal{Z} = \mathrm{Tr}_{\{\mathbf{s}_k\}}\exp{(-\beta\mathcal{H})}
=\mathrm{Tr}_{\{\mathbf{s}_k\}}\exp{\left(\frac{1}{2}\mathbf{s}_i\cdot(\beta\mathbb{I}_{ij})\cdot\mathbf{s}_j
+\mu\beta\mathbf{B}_i\cdot\mathbf{s}_i\right)},
\end{split}
\ee
where the Trace operation $\mathrm{Tr}_{\{\mathbf{s}_k\}}$ symbolises the summation over all spins $\mathbf{s}_k$ and their respective spin components. Using the following well-known identity for transforming Gaussian integrals \cite{papoulis2002probability}:
\be\label{hubb}
\begin{split}
\int_{-\infty}^{+\infty}&d\Omega
\exp{\left(-\frac{1}{2}\bm{\psi}_i\cdot(\beta^{-1}\mathbb{I}_{ij}^{-1})\cdot\bm{\psi}_j+\bm{\psi}_i\cdot\mathbf{s}_i\right)}
= |\beta\mathbb{I}|^{-1/2}\exp{\left(\frac{1}{2}\mathbf{s}_i\cdot(\beta\mathbb{I}_{ij})\cdot\mathbf{s}_j\right)}
\end{split}
\ee
allows to replace the interaction term in Eq. \eqref{pf0} and re-express the partition function as:
\be\label{pf1a}
\mathcal{Z}=
\frac{\beta^{3N/2}}{|\mathbb{I}|^{1/2}}\int_{-\infty}^{+\infty}d\Omega
\exp{(-\beta\mathcal{S})},
\ee
where:
\be\label{pf1b}
\begin{split}
\mathcal{S}=\frac{1}{2}\bm{\psi}_i&\cdot\mathbb{I}_{ij}^{-1}\cdot\bm{\psi}_j
-\frac{1}{\beta}\ln\mathrm{Tr}_{\{\mathbf{s}_k\}}\exp{(\beta(\bm{\psi}_i+\mu\mathbf{B}_i)\cdot\mathbf{s}_i)}
\end{split}
\ee
is the so-called Action. The new variables $\bm{\psi}_i$, $i=1,\dots,N$, represent continuous fields and act as conjugate variables to spins $\mathbf{s}_i$. In Eqs. \eqref{hubb}-\eqref{pf1b}, the integral volume element is $d\Omega = \prod_{i=1}^N{d\bm{\psi}_i}/(2\pi)^{3/2}$, the matrix $\mathbb{I}^{-1}$ denotes the inverse of the matrix $\mathbb{I}$, and $|\beta\mathbb{I}|^{-1/2} =\beta^{-3N/2}|\mathbb{I}|$ with $|\mathbb{I}|$ being the determinant of the matrix $\mathbb{I}$.

%%%%%%%%%%%%%%%%%%%%%%%%%%%%%%%%%%%%%%%%

\subsection{Magnetic moment and thermodynamic internal energy}

\noindent The thermal fluctuation-averaged magnetic moment at a site $i$ can be expressed by using the thermodynamic relation \cite{yeomans1992statistical}: 
\be\label{mi}
\mathbf{m}_i = -\frac{\partial\mathcal{F}}{\partial\mu\mathbf{B}_i}
= \frac{1}{\mu\beta}\frac{1}{\mathcal{Z}}\frac{\partial\mathcal{Z}}{\partial\mathbf{B}_i},
\ee
where $\mathcal{F}$ is the free energy related to the partition function $\mathcal{Z}$ as $\mathcal{F}=-\beta^{-1}\ln\mathcal{Z}$. Inserting Eq. \eqref{pf1a} in \eqref{mi} and arranging we obtain:
\be\label{miS}
\mathbf{m}_i = -\frac{1}{\mu}\frac
{\int_{-\infty}^{+\infty}d\Omega\frac{\partial\mathcal{S}}{\partial\mathbf{B}_i}\exp{(-\beta\mathcal{S})}}
{\int_{-\infty}^{+\infty}d\Omega\exp{(-\beta\mathcal{S})}},
\ee
which shows that the magnetic vector moment $\mathbf{m}_i$ at a given lattice site $i$ can be evaluated based on the knowledge of Action $\mathcal{S}$.
Similarly, the internal energy of the system can be obtained from the thermodynamic relation: 
\be\label{U}
\mathcal{U} = -\frac{\partial\ln\mathcal{Z}}{\partial\beta}
=-\frac{1}{\mathcal{Z}}\frac{\partial\mathcal{Z}}{\partial\beta}.
\ee
Inserting Eq. \eqref{pf1a} in \eqref{U} and arranging gives:
\be\label{US}
\mathcal{U} = -\frac{3N}{2\beta} + 
\frac
{\int_{-\infty}^{+\infty}d\Omega\left(\mathcal{S}+\beta\frac{\partial\mathcal{S}}{\partial\beta}\right)\exp{(-\beta\mathcal{S})}}
{\int_{-\infty}^{+\infty}d\Omega\exp{(-\beta\mathcal{S})}}.
\ee
The integral expressions in Eqs. \eqref{miS} and \eqref{US} will be evaluated below using the saddle-point approximation, which will allow obtaining the sought mean-field solution.\\

%%%%%%%%%%%%%%%%%%%%%%%%%%%%%%%%%%%%%%%%

\subsection{Mean-field model: saddle-point approximation}

\noindent The MF approximation is derived by considering that the main contribution to the integral in Eqs. \eqref{pf1a}, \eqref{miS} and \eqref{US} is given by the maximum of the exponential term, i.e. the minimum of Action $\mathcal{S}$. This minimum can be found from the extremum condition  $\partial\mathcal{S}/\partial\bm{\psi}_i=0$, evaluating which gives the following constraining relation between the continuous field variables:
\be\label{psimf}
\bm{\tilde\psi}_i = \sum_j\mathbb{I}_{ij}\frac
{\mathrm{Tr}_{\mathbf{s}_j}\mathbf{s}_j\exp{\left(\beta(\bm{\tilde\psi}_j+\mu\mathbf{B}_j)\cdot\mathbf{s}_j\right)}}
{\mathrm{Tr}_{\mathbf{s}_j}\exp{\left(\beta(\bm{\tilde\psi}_j+\mu\mathbf{B}_j)\cdot\mathbf{s}_j\right)}}.
\ee
The Trace operation is now taken only through the components of a single spin $\mathbf{s}_j$. Eq. \eqref{psimf} defines the continuous mean-fields $\bm{\tilde\psi}_i$. It represents a set of non-linear algebraic equations coupled through the interaction matrix $\mathbb{I}_{ij}$. Note the implicit temperature-, magnetic field- and interaction-dependence of the mean-fields $\bm{\tilde\psi}_i=\bm{\tilde\psi}_i(\beta, \mathbf{B}_j, \mathbb{I}_{ij})$, which was absent in the original $\bm{\psi}_i$. 
Expanding the Action (Eq. \eqref{pf1b}) around the mean-field solutions $\bm{\tilde\psi}_i$ gives:
\be\label{expansion}
\begin{split}
\mathcal{S} &=\mathcal{\tilde S}
+
\frac{1}{2}
\frac{\partial^2\mathcal{\tilde S}}{\partial\bm{\psi}_i\partial\bm{\psi}_j}\delta\bm{\psi}_i\delta\bm{\psi}_j+\dots,
\end{split}
\ee
where the fluctuations are $\delta\bm{\psi}_i = \bm{\psi}_i-\bm{\tilde\psi}_i$, $i=1,\dots,N$. The notation $\mathcal{\tilde S}$ and $\partial^2\mathcal{\tilde S}/\partial\bm{\psi}_i\partial\bm{\psi}_j$ implies, respectively, Eq. \eqref{pf1b} and its second partial derivative evaluated at the mean-fields $\bm{\tilde\psi}_i$. The missing first derivative term in Eq. \eqref{expansion} is zero due to the extremum condition requirement.
Inserting Eq. \eqref{expansion} in \eqref{miS} and arranging gives for the mean-field spin vector moment at a lattice site $i$:
\be\label{mimf_der}
\begin{split}
\mathbf{\tilde m}_i = -\frac{1}{\mu}\frac
{\int_{-\infty}^{+\infty}d\Omega\left(\frac{\partial\mathcal{\tilde S}}{\partial\mathbf{B}_i}+\dots\right)\mathrm{e}^{-\beta\mathcal{\tilde S}+\dots}}
{\int_{-\infty}^{+\infty}d\Omega\,\mathrm{e}^{-\beta\mathcal{\tilde S}+\dots}}
\approx-\frac{1}{\mu}\frac{\partial\mathcal{\tilde S}}{\partial\mathbf{B}_i}\frac
{\int_{-\infty}^{+\infty}d\Omega\,\mathrm{e}^{-\beta\mathcal{\tilde S}+\dots}}
{\int_{-\infty}^{+\infty}d\Omega\,\mathrm{e}^{-\beta\mathcal{\tilde S}+\dots}}
=-\frac{1}{\mu}\frac{\partial\mathcal{\tilde S}}{\partial\mathbf{B}_i},
\end{split}
\ee
where we neglected all higher order terms appearing inside the parentheses, brought the field-derivate term in front of the integral, and subsequently simplified the overall expression. Eq. \eqref{mimf_der} can be evaluated by first calculating the magnetic field derivative of Action in Eq. \eqref{pf1b}, and then replacing $\bm{\psi}_i$ by $\bm{\tilde\psi}_i$, which gives for the mean-field spin moments:
\be\label{mimf1}
\mathbf{\tilde m}_i = \frac
{\mathrm{Tr}_{\mathbf{s}_i}\mathbf{s}_i\exp{\left(\beta(\bm{\tilde\psi}_i+\mu\mathbf{B}_i)\cdot\mathbf{s}_i\right)}}
{\mathrm{Tr}_{\mathbf{s}_i}\exp{\left(\beta(\bm{\tilde\psi}_i+\mu\mathbf{B}_i)\cdot\mathbf{s}_i\right)}},
\ee
where the Trace operation is taken over the components of a spin $\mathbf{s}_i$. Comparing this result with Eq. \eqref{psimf} we immediately obtain the relation:
\be\label{psimi}
\bm{\tilde\psi}_i = \mathbb{I}_{ij}\mathbf{\tilde m}_j.
\ee
Inserting Eq. \eqref{psimi} in Eq. \eqref{mimf1} we obtain a more convenient form of the expression:
\be\label{mimf2}
\mathbf{\tilde m}_i = \frac
{\mathrm{Tr}_{\mathbf{s}_i}\mathbf{s}_i\exp{\left(\beta(\mathbb{I}_{ij}\mathbf{\tilde m}_j+\mu\mathbf{B}_i)\cdot\mathbf{s}_i\right)}}
{\mathrm{Tr}_{\mathbf{s}_i}\exp{\left(\beta(\mathbb{I}_{ij}\mathbf{\tilde m}_j+\mu\mathbf{B}_i)\cdot\mathbf{s}_i\right)}}.
\ee
Eq. \eqref{mimf2} represents a set of coupled non-linear equations for mean-field spin moments $\mathbf{\tilde m}_i$, which can be solved in a self-consistent manner as discussed in Sec. \ref{sec:itersol} in the main text.\\

\noindent Applying the procedure used to derive the key result of Eq. \eqref{mimf2} from \eqref{miS} now to Eq. \eqref{US} allows expressing the internal thermodynamic energy in the MF approximation as:
\be\label{USmf1}
\mathcal{\tilde U} = \mathcal{\tilde S}+\beta\frac{\partial\mathcal{\tilde S}}{\partial\beta},
\ee
where the additional factor $-3N/(2\beta)$ appearing in Eq. \eqref{US} has been absorbed into the definition of $\mathcal{\tilde U}$ since it is independent of the continuous field variables. Evaluating Eq. \eqref{USmf1} based on \eqref{pf1b}, and using Eqs. \eqref{mimf1} and \eqref{psimi} to express the continuous mean-field variables $\bm{\tilde\psi}_i$ in terms of the mean-field spin moments $\mathbf{\tilde m}_i$, we obtain the familiar expression:
\be\label{USmf2}
\mathcal{\tilde U} = - \frac{1}{2}\mathbf{\tilde m}_i\cdot\mathbb{I}_{ij}\cdot\mathbf{\tilde m}_j - \mu\mathbf{B}_i\cdot\mathbf{\tilde m}_i.
\ee
Rewriting this expression in the vector form used in Eq. \eqref{spinham} we obtain:
\be\label{USmf3}
\mathcal{\tilde U}
\equiv \mathcal{H_\mathrm{MF}} =
-\frac{1}{2}\sum_{ij}J_{ij}\mathbf{\tilde m}_i\cdot\mathbf{\tilde m}_j
-\frac{1}{2}\sum_{ij}\mathbf{D}_{ij}\cdot(\mathbf{\tilde m}_i\times\mathbf{\tilde m}_j)
-\mu\sum_i\mathbf{\tilde m}_i\cdot\mathbf{B}_i,
\ee
which we postulate as the mean-field ``Hamiltonian'', $\mathcal{H_\mathrm{MF}}$, consistent with the spin model defined by Eq. \eqref{spinham}. Differentiating this energy with respect to the mean-field spin moment gives:
\be\label{Beff1}
\begin{split}
\mu\mathbf{\tilde B}_i^\mathrm{e} 
= -\frac{\partial\mathcal{H_\mathrm{MF}}}{\partial\mathbf{\tilde m}_i}
=J_{ij}\mathbf{\tilde m}_j
- \mathbf{D}_{ij}\times\mathbf{\tilde m}_j
+ \mu\mathbf{B}_i.
\end{split}
\ee
Eq. \eqref{Beff1} represents the effective field acting at the lattice site $i$ due to the interaction with the neighbouring spins and the external magnetic field.
With this expression, Eq. \eqref{mimf2} can be rewritten in terms of the effective field:
\be\label{mimf3}
\mathbf{\tilde m}_i = \frac
{\mathrm{Tr}_{\mathbf{s}_i}\mathbf{s}_i\exp{\left(\beta\mu\mathbf{\tilde B}_i^\mathrm{e}\cdot\mathbf{s}_i\right)}}
{\mathrm{Tr}_{\mathbf{s}_i}\exp{\left(\beta\mu\mathbf{\tilde B}_i^\mathrm{e}\cdot\mathbf{s}_i\right)}},
\ee
which is our final expression for the mean-field moment of a spin $\mathbf{s}_i$. A computational method suitable for solving the systems of equations Eqs. \eqref{Beff1} and \eqref{mimf3} has been introduced in Sec. \ref{sec:itersol}, and is detailed in Appendix \ref{appendix:app2} below.

%%%%%%%%%%%%%%%%%%%%%%%%%%%%%%%%%%%%%%%%
%%%%%%%%%%%%%%%%%%%%%%%%%%%%%%%%%%%%%%%%
%%%%%%%%%%%%%%%%%%%%%%%%%%%%%%%%%%%%%%%%
%%%%%%%%%%%%%%%%%%%%%%%%%%%%%%%%%%%%%%%%

\section{Iterative solution of the mean-field model}\label{appendix:app2}

\noindent The Trace operation in Eq. \eqref{mimf3} can be written explicitly as a three-dimensional integral over the spin components, i.e. replacing $\mathrm{Tr}_{\mathbf{s}_i}\rightarrow\int d\mathbf{s}$:
\begin{subequations}\label{mimf4}
\be\label{mimf4a}
\mathbf{\tilde m}_i = \frac
{\int d\mathbf{s}\,\mathbf{s}\exp{\left(\beta\mu\mathbf{\tilde B}_i^\mathrm{e}\cdot\mathbf{s}\,\right)}}
{\int d\mathbf{s}\exp{\left(\beta\mu\mathbf{\tilde B}_i^\mathrm{e}\cdot\mathbf{s}\,\right)}}
\ee
\be
\mu\mathbf{\tilde B}_i^\mathrm{e} =
 J_{ij}\mathbf{\tilde m}_j
- \mathbf{D}_{ij}\times\mathbf{\tilde m}_j
+ \mu\mathbf{B}_i
\ee
\end{subequations}
An algorithm for solving the set of coupled non-linear equations Eqs. \eqref{mimf4} is based on the standard notion that the stable moment configurations correspond to the minima of the internal thermodynamic energy Eq. \eqref{USmf3}.\\

\noindent The stability conditions can be obtained by calculating the variation of the internal energy $\mathcal{U}\rightarrow\mathcal{U} + \delta\mathcal{U}$ resulting from an infinitesimal change of the magnetic moment $\mathbf{\tilde m}_i\rightarrow \mathbf{\tilde m}_i+\delta\mathbf{\tilde m}_i$. Evaluating the variation of energy Eq. \eqref{USmf3} and keeping only the terms up to the first order in $\delta\mathbf{\tilde m}_i$ gives $\delta\mathcal{U} = -\mathbf{\tilde B}_i^\mathrm{e}\cdot\delta\mathbf{\tilde m}_i$. The most general form of the rotational variation of a vector is $\delta\mathbf{\tilde m}_i = \mathbf{\tilde m}_i\times\delta\bm{\theta}$, which describes small rotation of $\mathbf{\tilde m}_i$ around an arbitrary axis identified by the direction of $\delta\bm{\theta}$. This leads to $\delta\mathcal{U} = -\mathbf{\tilde B}_i^\mathrm{e}\cdot(\mathbf{\tilde m}_i\times\delta\bm{\theta}) = (\mathbf{\tilde m}_i\times\mathbf{\tilde B}_i^\mathrm{e})\cdot\delta\bm{\theta}$. At an extremum, $\delta\mathcal{U} = 0$ for any arbitrary variation of $\delta\bm{\theta}$, which yields the state stability condition:
\be\label{stability}
\mathbf{\tilde m}_i\times\mathbf{\tilde B}_i^\mathrm{e} = 0
\hskip0.5cm
(\mathrm{or}
\hskip0.1cm
\mathbf{\tilde m}_i\parallel\mathbf{\tilde B}_i^e)
\hskip0.5cm
i=1,\dots, N.
\ee
The set of coupled equations Eq. \eqref{stability} implies that the stable configurations of moments $\mathbf{\tilde m}_i$, associated with the minima of the internal energy $\mathcal{U}$, are such that all $\mathbf{\tilde m}_i$ are aligned along their respective effective fields $\mathbf{\tilde B}_i^\mathrm{e}$. This allows to solve the integral in Eq. \eqref{mimf4a} as follows.\\

\noindent Multiplying (as dot product) both sites of Eq. \eqref{mimf4a} by the effective field $\mathbf{\tilde B}_i^\mathrm{e}$ gives:
\be\label{mimf5}
\mathbf{\tilde B}_i^\mathrm{e}\cdot\mathbf{\tilde m}_i = \frac
{\int d\mathbf{s}\,\mathbf{\tilde B}_i^\mathrm{e}\cdot\mathbf{s}\exp{\left(\beta\mu\mathbf{\tilde B}_i^\mathrm{e}\cdot\mathbf{s}\,\right)}}
{\int d\mathbf{s}\exp{\left(\beta\mu\mathbf{\tilde B}_i^\mathrm{e}\cdot\mathbf{s}\,\right)}}.
\ee
Since the stable states are such that $\mathbf{\tilde m}_i$ and $\mathbf{\tilde B}_i^\mathrm{e}$ are aligned, then $\mathbf{\tilde m}_i\cdot\mathbf{\tilde B}_i^\mathrm{e} = |\mathbf{\tilde m}_i||\mathbf{\tilde B}_i^\mathrm{e}|$ and $\mathbf{\tilde B}_i^\mathrm{e}\cdot\mathbf{s}=|\mathbf{\tilde B}_i^\mathrm{e}|\cos\phi$, where $\phi$ is the angle between the vectors $\mathbf{\tilde B}_i^\mathrm{e}$ and $\mathbf{s}$, and $|\mathbf{s}|=1$. Therefore, we can rewrite Eq. \eqref{mimf5} as:
\begin{equation}
|\mathbf{\tilde m}_i||\mathbf{\tilde B}_i^\mathrm{e}|
=
\frac{
\int d\mathbf{s}\,|\mathbf{\tilde B}_i^\mathrm{e}|\cos\phi\exp{
\left(\beta \mu|\mathbf{\tilde B}_i^\mathrm{e}|\cos\phi\right)}}{
\int d\mathbf{s}\,\exp{
\left(\beta \mu|\mathbf{\tilde B}_i^\mathrm{e}|\cos\phi\right)}},
\end{equation}
which upon dividing both sides of the equation by $|\mathbf{\tilde B}_i^\mathrm{e}|$ and integrating in spherical coordinates gives:
\be
|\mathbf{\tilde m}_i| = \mathcal{L}(\beta \mu|\mathbf{\tilde B}_i^\mathrm{e}|)
\ee
where $\mathcal{L}(x) = \coth(x) - {x}^{-1}$ is the Langevin function. This can be written in vector form using the requirement of alignment with the effective field given by Eq. \eqref{stability}:
\begin{subequations}\label{mimf6}
\be
\mathbf{\tilde m}_i = \mathcal{L}(\beta \mu|\mathbf{\tilde B}_i^\mathrm{e}|)
\frac{\mathbf{\tilde B}_i^\mathrm{e}}{|\mathbf{\tilde B}_i^\mathrm{e}|}
\ee
\be
\mu\mathbf{\tilde B}_i^\mathrm{e} =
 J_{ij}\mathbf{\tilde m}_j
- \mathbf{D}_{ij}\times\mathbf{\tilde m}_j
+ \mu\mathbf{B}_i
\ee
\end{subequations}
Thus the mean-field spin moment $\mathbf{\tilde m}_i$ can be evaluated directly based on the known form of a Langevin function and of the local effective field $\mathbf{\tilde B}_i^\mathrm{e}$. The explicit evaluation of the integrals in Eq. \eqref{mimf4} is not necessary. The set of coupled equations in Eq. \eqref{mimf6} can be solved using iterative technique highlighted in Sec. \ref{sec:itersol}.

\vfill
\end{document}